\documentclass[a4paper]{article}

\usepackage[margin=1in]{geometry}

\usepackage[T1]{fontenc}
\newcommand{\changefont}[3]{
\fontfamily{#1} \fontseries{#2} \fontshape{#3} \selectfont}

\changefont{ptm}{m}{n}

\usepackage{setspace} 

\usepackage{graphicx}    

\usepackage{amsfonts,amssymb,amsmath,amsgen,amsopn,amsbsy,theorem,graphicx,epsfig}
\usepackage{graphics}
\usepackage{mathrsfs}
\usepackage{color}
\newtheorem{remark}{Remark}[section]

\newtheorem{theorem}{Theorem}[section]

\newtheorem{definition}{Definition}[section]

\long\def\symbolfootnote[#1]#2{\begingroup%
\def\thefootnote{\fnsymbol{footnote}}\footnote[#1]{#2}\endgroup} 

\begin{document}

\begin{center}
\Large \textbf{A Novel Criterion for Unpredictable Motions}
\end{center}

\begin{center}
\normalsize \textbf{Fatma Tokmak Fen$^a$, Mehmet Onur Fen$^{b,}$\symbolfootnote[1]{Corresponding Author.  E-mail: monur.fen@gmail.com}, Marat Akhmet$^c$} \\
\vspace{0.2cm}
\textit{\textbf{\footnotesize$^a$Department of Mathematics, Gazi University, 06560 Ankara, Turkey}} \\
\textit{\textbf{\footnotesize$^b$Department of Mathematics, TED University, 06420 Ankara, Turkey}} \\
\textit{\textbf{\footnotesize$^c$Department of Mathematics, Middle East Technical University, 06800 Ankara, Turkey}}
\vspace{0.1cm}
\end{center}

\vspace{0.3cm}

\begin{center}
\textbf{Abstract}
\end{center}

\noindent\ignorespaces
We demonstrate the extension of unpredictable motions in coupled autonomous systems with skew product structure in the case that generalized synchronization takes place. Sufficient conditions for the existence of unpredictable motions in the dynamics of the response system are provided. The theoretical results are exemplified for coupled autonomous systems in which the drive is a  hybrid dynamical system and the response is a Lorenz system. The auxiliary system approach and conditional Lyapunov exponents are utilized to detect the presence of generalized synchronization. 

\vspace{0.2cm}
 
\noindent\ignorespaces \textbf{Keywords:} Unpredictable solution; Generalized synchronization; Auxiliary system approach; Conditional Lyapunov exponent; Poincar\'{e} chaos; Lorenz system 

\vspace{0.6cm}

\section{Introduction}

A special type of Poisson stable trajectory, named unpredictable, was introduced in paper \cite{Akhmetunpredictable}. An unpredictable trajectory leads to Poincar\'{e} chaos in the associated quasi-minimal set. Such trajectories take place in symbolic dynamics, logistic and H\'{e}non maps, and the Smale Horseshoe \cite{Akhmetunpredictable, Akhmetpoincare}. One of the important features of Poincar\'{e} chaos is that it can be triggered by the presence of a single unpredictable trajectory in the dynamics. This feature is the main difference of Poincar\'{e} chaos compared to chaos in the sense of Devaney \cite{Devaney87} and Li-Yorke \cite{Li75} since a collection of motions are required to define these chaos types.

Interesting results concerning unpredictable motions as well as Poincar\'{e} chaos in topological spaces were provided in papers \cite{Miller19}-\cite{Thakur21}. Miller \cite{Miller19} generalized the notion of unpredictable points to the case of semiflows with arbitrary acting abelian topological monoids, whereas Thakur and Das \cite{Thakur20} demonstrated that at least one of the factors is Poincar\'{e} chaotic provided that the same is true for finite or countably infinite products of semiflows. Additionally, differential equations with hyperbolic linear parts exhibiting unpredictable solutions were studied in \cite{Akhmetexistence, Akhmet17}, and the existence of unpredictable outputs in cellular neural networks can be found in papers \cite{Akhmet20,Fen21}.

In order to provide a larger class of differential equations possessing unpredictable trajectories, in this study we take into account coupled systems in which generalized synchronization (GS) \cite{Rulkov95} takes place. More precisely, we consider the  systems
\begin{eqnarray} \label{drivesyst1}
x'(t)=F(x(t)),
\end{eqnarray}
and
\begin{eqnarray} \label{respsyst2}
y'(t)=G(x(t), y(t)),
\end{eqnarray}
where $F:\mathbb R^p \to \mathbb R^p$ and $G:\mathbb R^p \times \mathbb R^q \to \mathbb R^q$ are continuous functions. Systems (\ref{drivesyst1}) and (\ref{respsyst2}) are respectively called the drive and response. It is worth noting that the coupled system (\ref{drivesyst1})-(\ref{respsyst2}) has a skew product structure. Our purpose is to rigorously prove that if the drive system (\ref{drivesyst1}) admits an unpredictable solution, then the same is true for the response system (\ref{respsyst2}) when they are synchronized in the generalized sense. Sufficient conditions to approve the unpredictability are given in Section \ref{sec3}.

The concept of synchronization in coupled chaotic systems was initiated by Pecora and Carroll \cite{Pecora90} for identical ones, and it is generalized for non-identical systems by Rulkov et al. \cite{Rulkov95}. GS characterizes the state of the response system when it is driven by the output of another system, the drive. It was proved by Kocarev and Parlitz \cite{Kocarev96} that GS occurs in coupled systems of the form (\ref{drivesyst1})-(\ref{respsyst2}) if and only if the response system (\ref{respsyst2}) is asymptotically stable for all initial values in a neighborhood of the chaotic attractor. When GS occurs a functional relation exists between the states of the drive and response systems \cite{Rulkov95,Kocarev96}. For that reason GS allows to predict the dynamics of the response system by the dynamics of the drive \cite{Miranda04,Abarbanel96}. Lyapunov exponent based conditions which imply that the response state is a smooth function of the drive state were provided by Hunt et al. \cite{Hunt97}.

In this study we require that the state of the response (\ref{respsyst2}) is a continuous function of the state of the drive (\ref{drivesyst1}). This criterion can be guaranteed by the auxiliary system approach, which was suggested by Abarbanel et al. \cite{Abarbanel96}. In this approach one needs to use another response system that is identical to the original one but independent of it, and monitor the final states after the transitories have died off when different initial data within the basin of attraction are utilized. GS occurs in the dynamics if the final states of the two response systems are identical \cite{Miranda04,Abarbanel96}. Mutual false nearest neighbors, conditional Lyapunov exponents, and Lyapunov functions are other techniques which can be used to determine GS \cite{Rulkov95,Kocarev96,Miranda04,He92}.

The novelty of the present study is the verification of unpredictability in coupled systems under the presence of GS. The concept of GS was not taken into account in the studies \cite{Akhmetpoincare,Akhmetexistence,Akhmet17}. For that reason a different proof technique is applied. The recent results are important not only from the theoretical point of view but also for applications since GS can occur in various real world problems concerning image encryption, secure communication, lasers, electronic circuits, and neural networks \cite{Moon21}-\cite{Huang06}.

\section{Preliminaries} \label{sec2}		
	
Throughout the paper we make use of the Euclidean norm for vectors. We suppose that	systems (\ref{drivesyst1}) and (\ref{respsyst2}) admit compact invariant sets $\Lambda_x \subset \mathbb R^p$ and $\Lambda_y \subset \mathbb R^q$, respectively. In this case, a solution of the coupled system (\ref{drivesyst1})-(\ref{respsyst2}) with initial condition from the set $\Lambda_x \times \Lambda_y$ remains in that set.

In what follows we consider a class of GS in which the state of the response (\ref{respsyst2}) is a continuous function of the state of the drive (\ref{drivesyst1}). More precisely, we say that GS occurs in the dynamics of the coupled system (\ref{drivesyst1})-(\ref{respsyst2}) if there is a continuous transformation $\psi$ such that for each $(x_0,y_0) \in \Lambda_x \times \Lambda_y$ the relation 
\begin{eqnarray*} \label{synchronization1}
\displaystyle \lim_{t\to\infty} \left\|y(t)-\psi(x(t))\right\|=0 
\end{eqnarray*} 
holds, where $x(t)$ and $y(t)$ are respectively the solutions of (\ref{drivesyst1}) and (\ref{respsyst2}) with $x(0)=x_0$ and $y(0)=y_0$. The reader is referred to the paper \cite{Abarbanel96} and the book \cite{Miranda04} for further information concerning continuity of the transformation $\psi$.

The definition of an unpredictable function which is utilized in the present study is as follows.

\begin{definition} (\cite{Akhmet17}) \label{unpfuncquasi}
A uniformly continuous function $h: \mathbb R \to \Lambda$, where $\Lambda$ is a compact subset of $\mathbb R^p$, is called unpredictable if there exist positive numbers $\epsilon_0$, $r$ and sequences $\left\{\mu_n\right\}_{n\in\mathbb N}$ and $\left\{\nu_n\right\}_{n\in\mathbb N}$ both of which diverge to infinity such that $\left\|h(t+\mu_n)-h(t)\right\| \to 0$ as $n \to \infty$ uniformly on compact subsets of $\mathbb R$ and $\left\|h(t+\mu_n)-h(t)\right\| \geq \epsilon_0$ for each $t \in [\nu_n-r,\nu_n+r]$ and $n \in \mathbb N.$
\end{definition}

The number $\epsilon_0$ in Definition \ref{unpfuncquasi} is called the unpredictability constant of the function $h(t)$ \cite{Akhmet17}.

\section{The Main Result}\label{sec3}

The following assumptions  on the response system (\ref{respsyst2}) are required.

\begin{itemize}
 \item[\textbf{(A1)}] There exists a positive number $L_1$ such that $\left\|G(x_1,y) - G(x_2,y) \right\| \geq L_1 \left\|x_1-x_2\right\|$ for each $x_1,x_2\in\Lambda_x$ and $y \in \Lambda_y$.
 \item[\textbf{(A2)}] There exists a positive number $L_2$ such that $\left\|G(x,y_1) - G(x,y_2) \right\| \leq L_2 \left\|y_1-y_2\right\|$ for each $x\in\Lambda_x$ and $y_1, y_2 \in \Lambda_y$.
\end{itemize}		
		
The following theorem provides a novel criterion for the existence of an unpredictable solution in the dynamics of the response system (\ref{respsyst2}).

\begin{theorem} \label{mainresultsynch}		
Suppose that the assumptions $(A1)$ and $(A2)$ are fulfilled. If the drive system (\ref{drivesyst1}) possesses an unpredictable solution and generalized synchronization takes place in the dynamics of the coupled system (\ref{drivesyst1})-(\ref{respsyst2}), then the response system (\ref{respsyst2}) also possesses an unpredictable solution.
\end{theorem}		

\noindent \textbf{Proof.}  
Let $x(t)$ be an unpredictable solution of the drive system (\ref{drivesyst1}). Since GS occurs in the dynamics of the coupled system (\ref{drivesyst1})-(\ref{respsyst2}), there is a continuous transformation $\psi$ and a point $y_0$ in $\Lambda_y$ such that the equation 
\begin{eqnarray} \label{synchronization}
\displaystyle \lim_{t\to\infty} \left\|y(t)-\psi(x(t))\right\|=0 
\end{eqnarray}  
is fulfilled, where $y(t)$ is the solution of the response system (\ref{respsyst2}) with $y(0)=y_0$. 

Let $\mathscr{C}$ be a compact subset of $\mathbb R$, and fix a positive number $\epsilon$. One can find real numbers $a$ and $b$ such that $\mathscr{C} \subseteq [a,b]$. According to (\ref{synchronization}) there is a positive number $T$ such that $$\left\|y(t) - \psi(x(t))\right\| < \displaystyle \frac{\epsilon}{3}$$ whenever $t \geq a+T$. 

It is worth noting that \begin{eqnarray} \label{zdefn} z(t)=(\widetilde{x}(t), \widetilde{y}(t)), \end{eqnarray} where $\widetilde{x}(t)=x(t+T)$ and $\widetilde{y}(t)=y(t+T)$, is a solution of the coupled system (\ref{drivesyst1})-(\ref{respsyst2}). In the proof we will show that $\widetilde{y}(t)$ is unpredictable. 

Because the transformation $\psi$ is continuous there is a positive number $\delta$ such that if $\left\|x_1-x_2\right\|<\delta$, then 
\begin{eqnarray} \label{proofineq11}
\left\|\psi(x_1) - \psi(x_2)\right\| < \displaystyle \frac{\epsilon}{3}.
\end{eqnarray}

Owing to the unpredictability of $x(t)$ there exist positive numbers $\epsilon_0$, $r$ and sequences $\left\{\mu_n\right\}_{n\in\mathbb N}$ and $\left\{\nu_n\right\}_{n\in\mathbb N}$ both of which diverge to infinity such that $\left\|x(t+\mu_n)-x(t)\right\| \to 0$ as $n \to \infty$ uniformly on compact subsets of $\mathbb R$ and $\left\|x(t+\mu_n)-x(t)\right\| \geq \epsilon_0$ for each $t \in [\nu_n-r,\nu_n+r]$ and $n \in \mathbb N$. Assume without loss of generality that $\mu_n \geq 0$ for each $n$. 

There is a natural number $n_0$ such that for $n \geq n_0$ and $t \in [a+T,b+T]$ the inequality 
$$\left\|x(t+\mu_n) - x(t) \right\|< \delta$$
holds. Hence, the inequality $$\left\| \psi(\widetilde{x}(t + \mu_n)) - \psi(\widetilde{x}(t))\right\| < \displaystyle \frac{\epsilon}{3}$$ is fulfilled for each $t \in [a,b]$ by means of (\ref{proofineq11}). Accordingly, for $n \geq n_0$ and $t \in \mathscr{C}$, it can be verified that
\begin{eqnarray*}
\left\|\widetilde{y}(t+\mu_n) - \widetilde{y}(t)\right\|  \leq  \left\| \widetilde{y}(t+\mu_n) - \psi(\widetilde{x}(t +\mu_n))\right\|
+ \left\| \psi(\widetilde{x}(t + \mu_n)) - \psi(\widetilde{x}(t))\right\| + \left\|\psi(\widetilde{x}(t)) - \widetilde{y}(t) \right\| 
 <\epsilon.
\end{eqnarray*}
For that reason, $\left\|\widetilde{y}(t+\mu_n)-\widetilde{y}(t)\right\| \to 0$ as $n \to \infty$ uniformly on compact subsets of $\mathbb R$.

In the remaining part of the proof, we will show the existence of numbers $\epsilon_1>0$, $r_1>0$ and a sequence $\left\{\theta_n \right\}$ with $\theta_n\to\infty$ as $n \to \infty$ such that $\left\|\widetilde{y}(t+\mu_n) - \widetilde{y}(t) \right\| \geq \epsilon_1$ for each $t \in [\theta_n -r_1, \theta_n+r_1]$ and $n \in \mathbb N$.

For each natural number $n$, we define $\omega_n = \nu_n-T$. Then, $$\left\|\widetilde{x}(t+\mu_n) - \widetilde{x}(t)\right\| \geq \epsilon_0$$ for each $t \in [\omega_n-r, \omega_n+r]$ and $n \in \mathbb N$. 

Next, let us denote $G(x,y)=(G_1(x,y),G_2(x,y),\ldots,G_q(x,y))$, where $G_i(x,y)$ is a real valued function for each $1 \leq i\leq q$.
Because the function $G(x,y)$ is uniformly continuous on the compact region $\Lambda_x \times \Lambda_y$ one can find a positive number $M_G$ such that $\left\|G(x,y)\right\| \leq M_G$ for each $x\in \Lambda_x, y \in \Lambda_y$. Since $\displaystyle \sup_{t \in \mathbb R}\left\|\widetilde{y}~'(t)\right\| \leq M_G$, the solution $\widetilde{y}(t)$ of the response system (\ref{respsyst2}) is uniformly continuous. Due to the uniform continuity of $\widetilde{x}(t)$, there exists a positive number $r_0$ such that
\begin{eqnarray} \label{proof1}
\left\|G(\widetilde{x}(t+\mu_n),\widetilde{y}(t))-G(\widetilde{x}(\mu_n+\omega_n),\widetilde{y}(\omega_n))\right\| \leq \displaystyle \frac{L_1 \epsilon_0}{4 \sqrt{q}}
\end{eqnarray}
and 
\begin{eqnarray} \label{proof2}
\left\|G(\widetilde{x}(t),\widetilde{y}(t))-G(\widetilde{x}(\omega_n),\widetilde{y}(\omega_n))\right\| \leq \displaystyle \frac{L_1 \epsilon_0}{4 \sqrt{q}}
\end{eqnarray}
for each $t \in [\omega_n - r_0, \omega_n+r_0]$ and $n \in \mathbb N$.

According to assumption $(A1)$, for each natural number $n$ there is an integer $i_n$ with $1 \leq i_n \leq q$ such that 
the inequality 
\begin{eqnarray} \label{proof3}
\left|G_{i_n}(\widetilde{x}(\mu_n+\omega_n),\widetilde{y}(\omega_n)) - G_{i_n} (\widetilde{x}(\omega_n),\widetilde{y}(\omega_n)) \right| \geq \displaystyle \frac{L_1}{\sqrt{q}} \left\|\widetilde{x}(\mu_n+\omega_n)-\widetilde{x}(\omega_n)\right\| \geq \displaystyle \frac{L_1 \epsilon_0}{\sqrt{q}}
\end{eqnarray}
is valid.

Fix a natural number $n$. For each $t \in [\omega_n-r_0, \omega_n+r_0]$, one can attain by means of the inequalities (\ref{proof1}), (\ref{proof2}), and (\ref{proof3}) that 
\begin{eqnarray*} \label{proof4}
\left|G_{i_n}(\widetilde{x}(t+\mu_n),\widetilde{y}(t)) - G_{i_n} (\widetilde{x}(t),\widetilde{y}(t)) \right| 
& \geq & \left|G_{i_n}(\widetilde{x}(\mu_n+\omega_n),\widetilde{y}(\omega_n)) - G_{i_n} (\widetilde{x}(\omega_n),\widetilde{y}(\omega_n)) \right| \\
&& - \left|G_{i_n}(\widetilde{x}(\mu_n+\omega_n),\widetilde{y}(\omega_n)) - G_{i_n} (\widetilde{x}(t+\mu_n),\widetilde{y}(t)) \right| \\
&& -  \left|G_{i_n}(\widetilde{x}(t),\widetilde{y}(t)) - G_{i_n} (\widetilde{x}(\omega_n),\widetilde{y}(\omega_n)) \right| \\
& \geq& \displaystyle \frac{L_1 \epsilon_0}{2\sqrt{q}}.
\end{eqnarray*}
There exist points $s_1, s_2, \ldots, s_q \in [\omega_n-r_0,\omega_n+r_0 ]$ such that 
\begin{eqnarray*}
\left\|\displaystyle \int_{\omega_n-r_0}^{\omega_n+r_0} \left[ G(\widetilde{x}(s+\mu_n),\widetilde{y}(s)) - G (\widetilde{x}(s),\widetilde{y}(s)) \right] ds \right\|
 & \geq  & 2r_0 \left|  G_{i_n}(\widetilde{x}(s_{i_n}+\mu_n),\widetilde{y}(s_{i_n})) - G_{i_n} (\widetilde{x}(s_{i_n}),\widetilde{y}(s_{i_n}))  \right| \\
& \geq & \displaystyle \frac{L_1 r_0 \epsilon_0}{\sqrt{q}}.
\end{eqnarray*}
Therefore,
\begin{eqnarray*}
\left\|\widetilde{y}(\mu_n+\omega_n+r_0)-\widetilde{y}(\omega_n+r_0)\right\|
& \geq & \left\|\displaystyle \int_{\omega_n-r_0}^{\omega_n+r_0} \left[ G(\widetilde{x}(s+\mu_n),\widetilde{y}(s)) - G (\widetilde{x}(s),\widetilde{y}(s)) \right] ds \right\| \\
&& - \left\| \widetilde{y}(\mu_n+\omega_n-r_0) -\widetilde{y}(\omega_n-r_0) \right\| \\
&& - \displaystyle \int_{\omega_n-r_0}^{\omega_n+r_0} \left\| G(\widetilde{x}(s+\mu_n),\widetilde{y}(s+\mu_n)) - G (\widetilde{x}(s+\mu_n),\widetilde{y}(s)) \right\| ds \\
&\geq & \displaystyle \frac{L_1 r_0 \epsilon_0}{\sqrt{q}} -  \left\| \widetilde{y}(\mu_n+\omega_n-r_0) -\widetilde{y}(\omega_n-r_0) \right\| \\
&& - \displaystyle \int_{\omega_n-r}^{\omega_n+r} L_2 \left\| \widetilde{y}(s+\mu_n) - \widetilde{y}(s)\right\| ds.
\end{eqnarray*}
The last inequality implies that
$$\displaystyle \max_{t \in [\omega_n-r_0, \omega_n+r_0]} \left\|\widetilde{y}(t+\mu_n) - \widetilde{y}(t)\right\| \geq \displaystyle \frac{L_1 r_0 \epsilon_0}{2 \sqrt{q} (r_0 L_2 +1)}. $$
Suppose that $$\displaystyle \max_{t \in [\omega_n-r_0, \omega_n+r_0]} \left\|\widetilde{y}(t+\mu_n) - \widetilde{y}(t)\right\| = \left\|\widetilde{y}(\mu_n+\zeta_n) -\widetilde{y}(\zeta_n)\right\| $$ for some $\zeta_n \in [\omega_n-r_0, \omega_n+r_0]$.
Let us denote $$\widetilde{r} = \displaystyle \frac{L_1 r_0 \epsilon_0}{8  M_G \sqrt{q}(r_0 L_2+1)}.$$
It can be confirmed for $t\in [\zeta_n-\widetilde{r}, \zeta_n+\widetilde{r}]$ that
\begin{eqnarray*}
\left\|\widetilde{y}(t+\mu_n) - \widetilde{y}(t)\right\| & \geq & \left\|\widetilde{y}(\mu_n+\zeta_n)-\widetilde{y}(\zeta_n)\right\| 
- \left| \displaystyle \int_{\zeta_n}^{t} \left\| G(\widetilde{x}(s+\mu_n), \widetilde{y}(s+\mu_n)) - G(\widetilde{x}(s), \widetilde{y}(s))  \right\|  ds \right| \\
& \geq & \displaystyle \frac{L_1 r_0 \epsilon_0}{2 \sqrt{q} (r_0 L_2 +1)} - 2 \widetilde{r} M_G \\
& = & \displaystyle \frac{L_1 r_0 \epsilon_0}{4 \sqrt{q} (r_0 L_2 +1)}.
\end{eqnarray*}
We define $\theta_n = \zeta_n + \widetilde{r}/2$ if $\zeta_n \in [\omega_n - r_0, \omega_n]$ and $\theta_n = \zeta_n - \widetilde{r}/2$ if $\zeta_n \in (\omega_n, \omega_n+r_0]$.

The inequality $\left\|\widetilde{y}(t+\mu_n) - \widetilde{y}(t)\right\| \geq \epsilon_1$ holds for each  $t \in [\theta_n-r_1,\theta_n+r_1]$ and $n \in \mathbb N$, where $$\displaystyle \epsilon_1 = \frac{L_1 r_0 \epsilon_0}{4 \sqrt{q} (r_0 L_2 +1)}$$ and $r_1 = \widetilde{r}/2$. The sequence $\left\{\theta_n\right\}_{n \in \mathbb N}$ diverges to infinity since the same is true for the sequence $\left\{\omega_n\right\}_{n \in \mathbb N}$. Thus, $\widetilde{y}(t)$ is unpredictable.
$\square$
			
\begin{remark} \label{remark1}
According to the proof of Theorem \ref{mainresultsynch}, for each $n \in \mathbb N$ the interval $[\theta_n-r_1,\theta_n+r_1]$ is a subset of the interval $[\omega_n-r, \omega_n+r]$. Therefore, the solution $z(t)=\left(\widetilde{x}(t), \widetilde{y}(t)\right)$ of the coupled system (\ref{drivesyst1})-(\ref{respsyst2})  defined by (\ref{zdefn}) is unpredictable such that $\left\|z(t+\mu_n)-z(t)\right\| \to 0$ as $n \to \infty$ uniformly on compact subsets of $\mathbb R$ and
$\left\|z(t+\mu_n)-z(t)\right\| \geq (\epsilon_0^2+\epsilon_1^2)^{1/2}$ for each $t \in [\theta_n-r_1,\theta_n+r_1]$ and $n \in \mathbb N$, in which $\epsilon_0$ and $\epsilon_1$ are respectively the unpredictability constants of $\widetilde{x}(t)$ and $\widetilde{y}(t)$. Hence, one can conclude under the assumptions of Theorem \ref{mainresultsynch} that the coupled system (\ref{drivesyst1})-(\ref{respsyst2}) possesses an unpredictable solution. 
\end{remark}

In the next section we demonstrate the extension of unpredictable solutions among coupled systems in which the drive is an autonomous hybrid system and the response is a Lorenz system.

\section{An Example}
	
According to the results of paper \cite{Akhmetpoincare}, the logistic map
\begin{eqnarray} \label{logistic}
\eta_{i+1}= \mu \eta_i (1-\eta_i)
\end{eqnarray}	
possesses an unpredictable orbit for the values of the parameter $\mu$ between $3+(2/3)^{1/2}$ and $4$. Moreover, for such values of $\mu$ the unit interval $[0,1]$ is invariant under the iterations of (\ref{logistic}) \cite{Hale91}.

Let $\left\{\eta^*_i\right\}_{i\in\mathbb Z}$ be an unpredictable orbit of (\ref{logistic}) with $\mu=3.94$, which belongs to the unit interval $[0,1]$, and suppose that $\gamma : \mathbb R \to [0,1]$ is the piecewise constant function defined by $\gamma(t)=\eta^*_i$ for $t \in (i, i+1]$, $i \in \mathbb Z$.
The function $\gamma(t)$ is the solution of the impulsive system
\begin{eqnarray} \label{gammafunc}
\gamma'(t)=0, \ \
\Delta \gamma |_{t=i}=\eta^*_{i}-\eta^*_{i-1} 
 \end{eqnarray}
satisfying the initial condition $\gamma(0)=\eta^*_{-1}$. The impulse moments of (\ref{gammafunc}) coincide with the ones of the solution of the discontinuous dynamical system 
\begin{eqnarray} \label{impmoments}
s'(t)=-1, \ \ 
\Delta s |_{s=0}=1 
\end{eqnarray}
with $s(0)=0$. 

It was demonstrated in study \cite{Akhmetexistence} that the differential equation
\begin{eqnarray} \label{difeqn}
\phi'(t)=- \phi(t) + \gamma(t),
\end{eqnarray}
admits the unique uniformly continuous unpredictable solution
\begin{eqnarray*} \label{unpfunc}
\phi(t)= \displaystyle \int_{-\infty}^t e^{-(t-s)} \gamma(s) ds,
\end{eqnarray*}
which is globally exponentially stable. Theorem $5.2$ \cite{Akhmetpoincare}, on the other hand, implies that the function $$\widetilde{\phi}(t)=\big( 2 \phi(t)+ 0.1\sin(\phi(t)), \ 3 \phi(t), \ 2.5\phi(t) + 0.2\cos(\phi(t)) \big)$$ is also unpredictable. For that reason, the system		
\begin{eqnarray} \label{system1}
&& x'_1(t)=-0.5x_1(t) + 2 \phi(t)+ 0.1\sin(\phi(t)) \nonumber \\
&& x'_2(t)=-0.3 x_2(t) + 0.2 \arctan(x_1(t)) + 3 \phi(t)  \\
&& x'_3(t)=-0.4x_3(t)+2.5\phi(t) + 0.2\cos(\phi(t)) \nonumber
\end{eqnarray}			
possesses a unique unpredictable solution by Theorem 4.1  \cite{Akhmetexistence}. Accordingly, the autonomous hybrid system (\ref{logistic})-(\ref{gammafunc})-(\ref{impmoments})-(\ref{difeqn})-(\ref{system1}) has a unique unpredictable solution. 

\begin{figure}[ht!] 
\centering
\includegraphics[width=14.0cm]{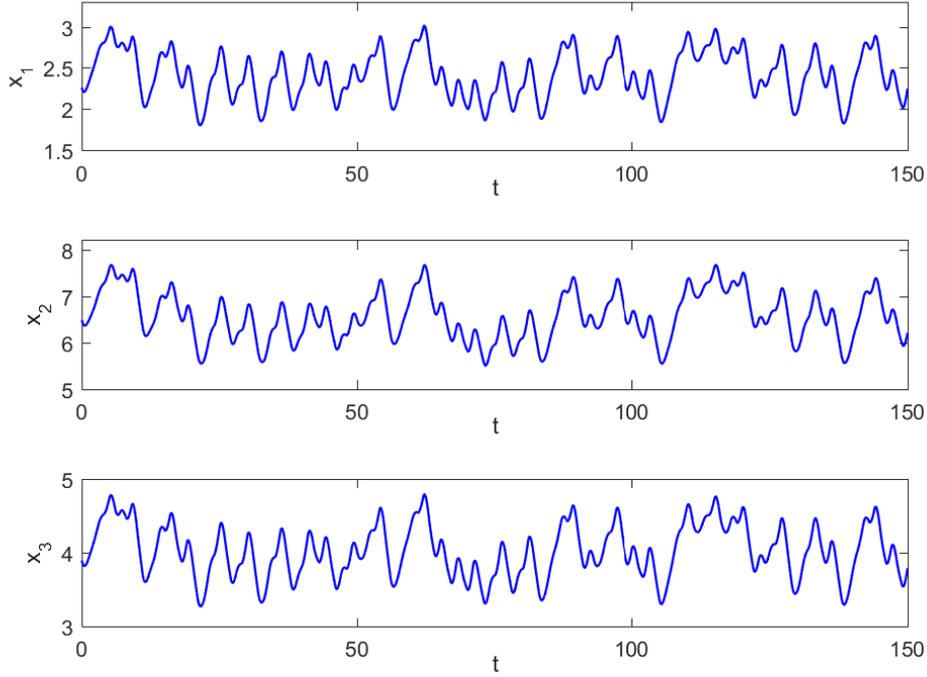}
\caption{Time-series of the $x_1$, $x_2$, and $x_3$-coordinates of the autonomous hybrid system (\ref{logistic})-(\ref{gammafunc})-(\ref{impmoments})-(\ref{difeqn})-(\ref{system1}). Each time-series is irregular, and this confirms the presence of an unpredictable solution in the dynamics.}
\label{fig1}
\end{figure}

Figure \ref{fig1} shows the time-series of the $x_1$, $x_2$, and $x_3$-coordinates of the solution of this hybrid system corresponding to the initial data $\zeta_0=0.76$, $\phi(0)=0.41$, $x_1(0)=2.26$, $x_2(0)=6.48$, $x_3(0)=3.89$, and value of the parameter $\mu=3.94$. The irregularity of each of the time-series confirms the presence of an unpredictable solution.

Next, let us consider the Lorenz system \cite{Lorenz63}
\begin{eqnarray} \label{system2}
&& y'_1(t)=-20y_1(t)+20y_2(t) \nonumber \\
&& y'_2(t)=-y_1(t) y_3(t)+41.05 y_1(t)-y_2(t) \\
&& y'_3(t)=y_1(t) y_2(t)-3 y_3(t). \nonumber 
\end{eqnarray}
We use system (\ref{system1}) as the drive, and establish unidirectional coupling between (\ref{system1}) and (\ref{system2}) by setting up the response system
\begin{eqnarray} \label{system3}
&& y'_1(t)=-20y_1(t)+20y_2(t)+2.9 x_1 (t) \nonumber \\
&& y'_2(t)=-y_1(t) y_3(t)+41.05 y_1(t)-y_2(t)+2.6 x_2(t) \\
&& y'_3(t)=y_1(t) y_2(t)-3 y_3(t)+2.4 x_3(t), \nonumber
\end{eqnarray}
where $x(t)=(x_1(t), x_2(t), x_3(t))$ is a solution of (\ref{system1}). It can be verified that the assumptions $(A1)$ and $(A2)$ hold for system (\ref{system3}).

\begin{figure}[ht!] 
\centering
\includegraphics[width=11.0cm]{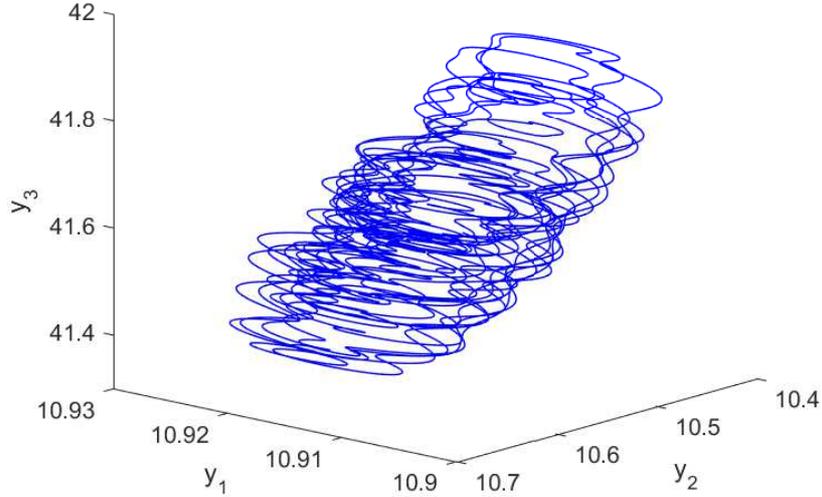}
\caption{The projection of the trajectory of autonomous hybrid system (\ref{logistic})-(\ref{gammafunc})-(\ref{impmoments})-(\ref{difeqn})-(\ref{system1})-(\ref{system3}) on the $y_1-y_2-y_3$ space. The irregularity observed in the figure approves that system (\ref{system3}) admits an unpredictable solution.}
\label{fig2}
\end{figure}

Using the solution of the drive system (\ref{system1}) whose coordinates are depicted in Figure \ref{fig1}, the trajectory of the response system (\ref{system3}) shown in Figure \ref{fig2} is obtained. In the simulation, the initial data $y_1(0)=10.91$, $y_2(0)=10.58$, and $y_3(0)=41.62$ are utilized. The irregular behavior of this trajectory reveals the presence of an unpredictable solution. 

\begin{figure}[ht!] 
\centering
\includegraphics[width=9.0cm]{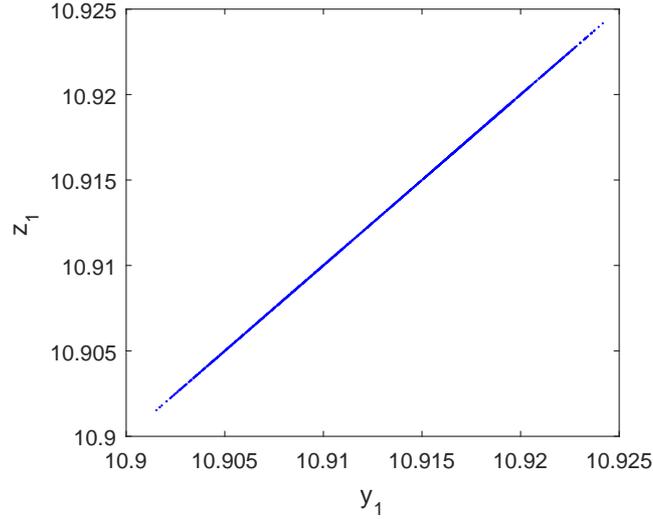}
\caption{The result of the auxiliary system approach applied to the coupled system (\ref{system1})-(\ref{system3}). The figure confirms that (\ref{system1}) and (\ref{system3}) are synchronized and there is an unpredictable solution of the response system (\ref{system3}).}
\label{fig3}
\end{figure}

Now, we will make use of the auxiliary system approach \cite{Abarbanel96} to show the presence of GS. We take into account the auxiliary system
\begin{eqnarray} \label{auxiliary}
&& z'_1(t)=-20z_1(t)+20z_2(t)+2.9 x_1 (t) \nonumber \\
&& z'_2(t)=-z_1(t) z_3(t)+41.05 z_1(t)-z_2(t)+2.6 x_2(t) \\
&& z'_3(t)=z_1(t) z_2(t)-3 z_3(t)+2.4 x_3(t), \nonumber
\end{eqnarray}
which is an identical copy of the response system (\ref{system3}). Again using the solution of (\ref{system1}) shown in Figure \ref{fig1} and the initial data $y_1(0)=10.91$, $y_2(0)=10.58$, $y_3(0)=41.62$, $z_1(0)=10.52$, $z_2(0)=10.03$, and $z_3(0)=41.96$, we represent in Figure \ref{fig3} the projection of the stroboscopic plot of the hybrid system (\ref{logistic})-(\ref{gammafunc})-(\ref{impmoments})-(\ref{difeqn})-(\ref{system1})-(\ref{system3})-(\ref{auxiliary})
on the $y_1-z_1$ plane. The plot is obtained by omitting the first $200$ iterations in order to eliminate the transients. Because it takes place on the line $z_1=y_1$, one can confirm that GS occurs, and therefore, there exists a continuous transformation $\psi$, which has no explicit time dependence, such that the equation 
$$
\displaystyle \lim_{t \to \infty} \left\|y(t) - \psi(x(t))\right\|=0
$$ 
is fulfilled, where $x(t)=(x_1(t), x_2(t), x_3(t))$ and $y(t)=(y_1(t), y_2(t), y_3(t))$ are respectively the states of (\ref{system1}) and (\ref{system3}) \cite{Miranda04,Abarbanel96}.
Hence, in accordance with our theoretical results, the response system (\ref{system3}) admits an unpredictable solution. Moreover, an unpredictable motion takes place also in the dynamics of the coupled system (\ref{system1})-(\ref{system3}) in accordance with Remark \ref{remark1}.  

To approve the presence of GS one more time, let us evaluate the conditional Lyapunov exponents of the response (\ref{system3}). For that purpose, we take into account the corresponding variational system
\begin{eqnarray} \label{variational}
&& w'_1(t)=-20w_1(t)+20w_2(t) \nonumber \\
&& w'_2(t)=(-y_3(t)+41.05) w_1(t)-w_2(t)-y_1(t) w_3(t) \\
&& w'_3(t)= y_2(t)w_1(t)+y_1(t)w_2(t)-3 w_3(t). \nonumber
\end{eqnarray}
When the solution $y(t)=(y_1(t), y_2(t), y_3(t))$ of (\ref{system3}) with $\zeta_0= 0.76$, $\phi(0)=0.41$, $x_1(0)= 2.26$, $x_2(0)=   6.48$, $x_3(0)= 3.89$, $y_1(0)= 10.91$, $y_2(0)= 10.58$, $y_3(0)= 41.62$, and $\mu=3.94$ is utilized, the largest Lyapunov exponent of system (\ref{variational}) is calculated as $-0.2371$, i.e., all conditional Lyapunov exponents of system (\ref{system3}) are negative. This confirms that (\ref{system1}) and (\ref{system3}) are synchronized in generalized sense \cite{Kocarev96,Miranda04}, and accordingly, the response system (\ref{system3}) as well as the coupled system (\ref{system1})-(\ref{system3}) possess unpredictable solutions.

\section{Concluding Remarks}

Synchronization is one of the phenomena that can occur in coupled chaotic systems \cite{Miranda04}. This phenomenon can be observed in various fields such as image encryption, secure communication, lasers, electronic circuits, and neural networks \cite{Moon21}-\cite{Huang06}. The presence of an unpredictable trajectory makes the corresponding dynamics exhibit chaotic behavior \cite{Akhmetunpredictable}. In this paper it is rigorously proved that if a drive system possesses an unpredictable solution, then the response also has the same property under the conditions mentioned in Section \ref{sec3}.  
The proposed technique makes it possible to obtain high dimensional systems possessing unpredictable trajectories.
In the future, our approach can be used to detect unpredictable trajectories in time-delayed unidirectionally or mutually coupled time-delayed systems \cite{Moskalenko21}.

\section*{Acknowledgments}

M. Akhmet has been supported by 2247-A National Leading Researchers Program of T\"{U}B\.{I}TAK, No. 120C138.


\begin{thebibliography}{99}
	
\bibitem{Akhmetunpredictable} M. Akhmet, M. O. Fen,  Unpredictable points and chaos,  Commun. Nonlinear Sci. Numer. Simulat. \textbf{40} (2016) 1-5.

\bibitem{Akhmetpoincare} M. Akhmet, M. O. Fen, Poincar\'{e} chaos and unpredictable functions,  Commun. Nonlinear Sci. Numer. Simulat. \textbf{48} (2016) 85-94. 

\bibitem{Devaney87} R. Devaney, An Introduction to Chaotic Dynamical Systems, United States of America, Addison-Wesley, 1987.

\bibitem{Li75} T. Y. Li, J. A. Yorke, Period three implies chaos, Am. Math. Mon. \textbf{82} (1975) 985-992.	

\bibitem{Miller19} A. Miller, Unpredictable points and stronger versions of Ruelle-Takens and Auslander-Yorke chaos, Topol. Appl. \textbf{253} (2019) 7-16.  

\bibitem{Thakur20} R. Thakur, R. Das, Strongly Ruelle-Takens, strongly Auslander-Yorke and Poincar\'e chaos on semiflows, Commun. Nonlinear Sci. Numer. Simulat. \textbf{81} (2020) 105018. 

\bibitem{Thakur21} R. Thakur, R. Das, Sensitivity and chaos on product and on hyperspatial semiflows, J. Differ. Equ. Appl. \textbf{27} (2021) 1-15. 

\bibitem{Akhmetexistence} M. Akhmet, M. O. Fen,  Existence of unpredictable solutions and chaos, Turk. J. Math. \textbf{41} (2017) 254-266.

\bibitem{Akhmet17} M. Akhmet, M. O. Fen, Non-autonomous equations with unpredictable solutions, Commun. Nonlinear Sci. Numer. Simulat. \textbf{59} (2017) 657-670.

\bibitem{Akhmet20} M. Akhmet, R. Seilova, M. Tleubergenova, A. Zhamanshin, Shunting inhibitory cellular neural networks with strongly unpredictable oscillations, Commun. Nonlinear Sci. Numer. Simulat. \textbf{89} (2020) 105287.

\bibitem{Fen21} M. O. Fen, F. Tokmak Fen, Unpredictable oscillations of SICNNs with delay, Neurocomputing \textbf{464} (2021) 119-129.

\bibitem{Rulkov95}  N. F. Rulkov, M. M. Sushchik, L. S. Tsimring, H. D. I. Abarbanel, Generalized synchronization of chaos in directionally coupled chaotic systems, Phys. Rev. E \textbf{51} (1995) 980-994.	
	
\bibitem{Pecora90} L. M. Pecora, T. L. Carroll, Synchronization in chaotic systems, Phys. Rev. Lett. \textbf{64} (1990) 821-825. 
	
\bibitem{Kocarev96} L. Kocarev, U. Parlitz, Generalized synchronization, predictability, and equivalence of unidirectionally coupled dynamical systems, Phys. Rev. Lett. \textbf{76} (1996) 1816-1819.		
	
\bibitem{Miranda04} J. M. Gonz\'{a}les-Miranda, Synchronization and Control of Chaos, Imperial College Press, London, 2004.
			
\bibitem{Abarbanel96} H. D. I. Abarbanel, N. F. Rulkov, M. M. Sushchik, Generalized synchronization of chaos: the auxiliary system approach, Phys. Rev. E \textbf{53} (1996)	4528-4535.

\bibitem{Hunt97} B. R. Hunt, E. Ott, J. A. Yorke, Differentiable generalized synchronization of chaos, Phys. Rev. E \textbf{55} (1997) 4029-4034.

\bibitem{He92} R. He, P. G. Vaidya, Analysis and synthesis of synchronous periodic and chaotic systems, Phys. Rev. A \textbf{46} (1992) 7387-7392.
	
\bibitem{Moon21} S. Moon, J.-J. Baik, J. M. Seo, Chaos synchronization in generalized Lorenz and an application to image encryption, Commun. Nonlinear Sci. Numer. Simulat. \textbf{96} (2021) 105708. 

\bibitem{Kinzel10} W. Kinzel, A. Englert, I. Kanter, On chaos synchronization and secure communication, Phil. Trans. R. Soc. A \textbf{368} (2010) 379-389.	

\bibitem{Uchida03} A. Uchida, K. Higa, T. Shiba, S. Yoshimori, F. Kuwashima, H. Iwasawa, Generalized synchronization of chaos in He-Ne lasers, Phys. Rev. E. \textbf{68} (2003) 016215	

\bibitem{Silva06} I. G. D. Silva, J. M. Buld\'{u}, C. R. Mirasso, J. Garc\'{i}a-Ojalvo, Synchronization by dynamical relaying in electronic circuit arrays, Chaos \textbf{16} (2006) 043113.

\bibitem{Huang06} X. Huang, J. Cao, Generalized synchronization for delayed chaotic neural networks: a novel coupling scheme, Nonlinearity \textbf{19} (2006) 2797-2811.

\bibitem{Hale91} J. Hale, H. Ko\c{c}ak, Dynamics and Bifurcations, New York, Springer-Verlag, 1991.

\bibitem{Lorenz63} E. N. Lorenz, Deterministic nonperiodic flow, J. Atmos. Sci. \textbf{20} (1963) 130-141.
		
\bibitem{Moskalenko21} O. I. Moskalenko, A. A. Koronovskii, A. D. Plotnikova, Pecularities of generalized synchronization in unidirectionally and mutually coupled time-delayed systems, Chaos Solit. Fract. \textbf{148} (2021) 111031.
	
\end{thebibliography}
\end{document}